# Galaxy rotation curves with log-normal density distribution

John H Marr

*Unit of Computational Science, Building 250, Babraham Research Campus, Cambridge, CB22 3AT, UK.*

E-mail: john.marr@2from.com

**ABSTRACT**

The log-normal distribution represents the probability of finding randomly distributed particles in a micro canonical ensemble with high entropy. To a first approximation, a modified form of this distribution with a truncated termination may represent an isolated galactic disk, and this disk density distribution model was therefore run to give the best fit to the observational rotation curves for 37 representative galaxies. The resultant curves closely matched the observational data for a wide range of velocity profiles and galaxy types with rising, flat or descending curves in agreement with Verheijen's classification of 'R', 'F' and 'D' type curves, and the corresponding theoretical total disk masses could be fitted to a baryonic Tully Fisher relation (bTFR). Nine of the galaxies were matched to galaxies with previously published masses, suggesting a mean excess dynamic disk mass of dex$0.61 \pm 0.26$ over the baryonic masses. Although questionable with regard to other measurements of the shape of disk galaxy gravitational potentials, this model can accommodate a scenario in which the gravitational mass distribution, as measured via the rotation curve, is confined to a thin plane without requiring a dark-matter halo or the use of MOND.

## 1. INTRODUCTION

The observations of flat rotation curves (RCs) in disk galaxies first reported by Rubin have created a number of problems in their interpretation. This has led many observers to postulate the existence of a dark matter (DM) halo, whose properties combine additively with baryonic matter to produce flat RCs (Casertano & van Gorkom 1991). A number of DM candidates have some theoretical justification and their existence may explain other observational data, but the inability of experimentalists to discover any evidence for suitable DM particles has led other observers to postulate that Newtonian gravity is incomplete, with the gravitational constant varying at weak field strengths to produce the observed RCs (the MOND hypothesis, Milgrom 1983). Advocates of MOND, including de Blok & McGaugh (1998), Sanders (1999), Sanders & McGaugh (2002), and Swaters, Sanders & McGaugh (2010), support the idea that gravitational dynamics is non-Newtonian in the limit of low accelerations and that it is unnecessary to invoke the presence of large quantities of unseen matter. Although MOND can be adjusted to fit the observed curves well, this ad hoc adjustment lacks theoretical justification at the present time. Other attempts to describe galactic dynamics without recourse to dark matter include the application of a modified acceleration law obtained from Einstein gravity coupled to a massive skew-symmetric field (Brownstein & Moffat 2006); general relativistic attempts to explain flat galactic RCs (Balasin & Grumiller 2006); and a logarithmic correction to the Newtonian field (Fabris & Pereira Campos 2009).

The gravitational potential and resultant motion of a point test mass within a model galaxy is a complex function of the mass distribution within the galaxy. Unlike the gravitational potential within and external to a uniform massive sphere, that within a disk of matter does not yield to simple analytical methods of integration but requires numerical methods for its solution. Historically, a number of papers have discussed the analysis of RCs of thin disk galaxies as a function of the surface density. Eckhardt & Pestaňa (2002) presented a technique for calculating the midplane gravitational potential of a thin axisymmetric galactic disk. They derived a number of mathematical expressions for assessing the compatibility of observed brightness and Doppler distributions of galactic disks, and for testing a number of gravitational theories, using Wolfram Mathematica™ to compute the derived functions. Simple methods of numerical integration have been presented by Nicholson (2003) and Banhatti (2008), while Kochanek (2008) has described an integration method for equatorial RCs. More recently Keeports (2010) has demonstrated the construction and evaluation of an integral for the gravitational field for an idealized planar galaxy with circular symmetry as a function of radial distance, and Jalocha et al (2010) described a global disk model that matched observed luminosity curves of 5 galaxies while giving a good fit to their rotation curves.

Analytical techniques suffer from four major problems: (1) The mathematical analysis is complex, and they can only be used with simple assumptions about the surface density distribution; (2) They often suffer from an infinity problem as the radial integration crosses the radial position of the test mass, sometimes solved by placing the test mass off centre from the galactic disk; (3) They generally have to extend the galactic radius to infinity, which demands an exponentially decreasing surface density; (4) They are unable to cope with a finite boundary. By using numerical integration rather than an analytical technique, it is relatively easy to derive these curves for any density distribution, including boundary conditions at a finite $R_{max}$, and any arbitrary gravitational law. The infinity problem is also easily solved by programmatically handling division by zero errors.

The inverse problem – the derivation of galactic disk surface density profiles from the observed rotation curves – is computationally difficult. Given the velocity curve, Toomre (1963) derived a theoretical function to obtain the mass density distribution to produce the curve based on the availability of appropriate mathematical analytical functions. Toomre stressed that no unique advantages could be claimed for these models except that both their rotation and density laws could be exactly expressible in terms of relatively simple functions, although the oscillatory behaviour of the Bessel functions made the integration difficult except for special solution families. This approach was further developed by Freeman (1970), Kent (1986), Cuddeford (1993), and Conway (2000). In contrast, Jalocha et al (2010) suggested building a catalogue of simulated velocity fields using N-body simulations to include observed qualitative characteristics such as the number of arms or the presence of a bar as a better way towards realistic modelling of the mass distribution in galaxies.

A different approach taken for this paper used iterative feedback to modulate the density profile of a model disk until the numerically calculated RC matched the observational curves for a small number of typical galaxies (NGC 2915, NGC 3521,F563-V2), assuming a thin flat axisymmetric disk, with no bulge or halo. The resultant density curves had the appearance of a truncated log-normal density distribution function, and this function was therefore run against a number of galaxies using a curve-fitting algorithm to generate the parameters for each velocity profile. The model curves produced with this function closely matched the observational data for a wide range of velocity profiles and galaxy types, and the calculated theoretical masses also fitted a baryonic Tully-Fisher relation (bTFR) reasonably well.

## 2. MODELLING ROTATION CURVES WITH A LOG-NORMAL DENSITY DISTRIBUTION

We may generally consider a thin disk to be a series of concentric annuli of equal width $\delta r$, radii $r$, and negligible thickness $\delta h$, and we may legitimately ask, what is the probability that any given star will be found in an area of disk $\delta A$ in the annulus at $r$? Knowing nothing of the star's history we may only state that this probability will be some unknown function of a number of independent variables such as its initial position, momentum, and the mass distribution of the system. The probability distribution for such a system of products of variables, which is also the typical distribution for a maximum entropy system, is the log-normal distribution (Limpert et al. 2001) with the general form:

$$p_x(x; \mu, \sigma) = \frac{1}{x\sigma\sqrt{2\pi}} \exp\left(-\frac{(\log x - \mu)^2}{2\sigma^2}\right), \ x > 0 \quad (1)$$

where $p_x \, dx$ is the probability that the variable will be found between $x$ and $x + dx$. For the galactic disk, this becomes the probability that any individual disk star will be found in an area $\delta A$ in the annulus at $r$, and the log-normal distribution may be modified to:

$$\Sigma(r) = \frac{\Sigma_0}{r/r_\mu} \exp\left(-\frac{[\log(r/r_\mu)]^2}{2\sigma^2}\right), \; r > 0 \qquad (2)$$

where $\Sigma(r)$ is the disk surface density ($M_\odot$ kpc$^{-2}$), $r$ is radial distance (kpc), $r_\mu$ is the mean of the natural logarithm of the radius (kpc), $\sigma$ is the standard deviation of the natural logarithm of the radius, and $\Sigma_0$ is a surface density parameter ($M_\odot$ kpc$^{-2}$). Strictly, it should be noted that the function applies to the distribution of all dynamical (gravitational) mass and will therefore include the total contribution of baryonic mass (stars, gas clouds, HI, dust, etcetera) and any gravitationally bound DM. Because Equation (2) gives an integrable function for the annular mass, the total disk mass can be calculated from:

$$M_{disk} = \Sigma_0 \sqrt{2\pi^3} r_\mu^2 \sigma \exp\left(\frac{\sigma^2}{2}\right)\left[1 - \mathrm{erf}\left(\frac{\sigma^2 - \ln(R_{max}/r_\mu)}{\sigma\sqrt{2}}\right)\right]$$
(3)

where $R_{max}$ is a maximum radius for the disk (kpc).

### 2.1. Rationale for a log-normal probability distribution of mass in a thin galactic disk

The rationale for selecting a log-normal probability (or Galton distribution) for the rotating mass in a galactic disk is supported on several grounds. First, disk galaxies are expected to have had a rather quiescent dynamical evolution since z~0.6. Puech, Hammer & Flores et al. (2009) looked at stellar mass-TFR (smTFR) at distances out to z~0.6, and suggested that the large scatter in the smTRF at that distance was due to major mergers up to that epoch, but there has been an absence of evolution in the bTFR over the past 6 Gyr, implying that no external gas accretion is required for distant rotating disks to sustain star formation. Gurovich et al. (2010) similarly concluded that the total baryon content of isolated disk galaxies (as measured by stellar+1.4 Hi mass) has not been much affected by galaxy evolution, and bTFR might be a fundamental relation back to the main epoch of galaxy assembly.

An important consequence is that several independent physical parameters of the system are conserved such as the total mass, the total angular momentum, and the total internal energy as a summation of kinetic energy and gravitational potential energy. Being isolated from any external transfer of energy or mass, this is defined thermodynamically as a micro canonical ensemble with its initial parameters fixed and invariant. An isolated galaxy may therefore be considered, for much of its existence, as an independent collisionless assemblage of discrete particles (stars), interacting with each other gravitationally. The probability distribution function for such a system is the log-normal distribution, the characteristics of which are fulfilled by the density distribution of spiral galaxies: the radial variable can never be negative, and the distribution cannot be Gaussian, but must be highly skewed (Feigelson & Babu 2012). The vast number of stars involved justifies treating their distribution as a continuum; the radius where the star might be found must be > 0 and can extend to large $r$; and the normalisation of the function reflects the certainty that the total probability of an individual star being somewhere in the disk must be 1 (this only applies exactly if $r\to\infty$, but in practice the probability function approaches zero rapidly for large $r$). The fact that $\Sigma(r)\to 0$ as $r\to 0$ is also reasonable, as the function applies only to disk stars, while the collapse of the rotation curve near the galactic centre reflects the region where bulge stars predominate.

### 2.2. The gravitational potential within a thin massive disk

A thin disk may be considered in the limit as an infinite number of annuli of width d$r$. Letting the disk have radius $R_{max}$, the total radial gravitational force/unit mass at a distance $r_i$ from the centre within the plane of the disk is given by the double integral:

$$F_r(r_i) = G \int_0^{Rmax} \int_0^{2\pi} \frac{\Sigma(r)\, r(r_i - r\cos\alpha)}{(r^2 + r_i^2 - 2rr_i\cos\alpha)^{3/2}} \, d\alpha\, dr \qquad (4)$$

where $\alpha$ is the angle subtended from $m_i$ through the centre to a point on the annulus at $r$, $F_r$ is directed towards the centre, and $\Sigma(r)$ is defined as the surface density at radius $r$ (Keeports 2010). For the log-normal distribution models, $\Sigma(r)$ was substituted from Equation (2). The analytical integral solution is non-trivial, containing components of elliptical integrals of the first and second kind, but this may be solved by numerical integration using standard tools such as ROTMOD in the GIPSY software, derived from work by Casertano (1983). The linear velocity $V(r_i)$ for the unit mass moving in a stable circular orbit at equilibrium may be calculated using the relationship:

$$V(r_i) = \sqrt{F_r(r_i) \cdot r} \qquad (5)$$

### 2.3. Fitting the model to the data

A typical rotation curve is shown in Figure 1 for F563-V2, with the fitted log-normal model curve overlain (solid line). $R_{max}$ was taken to be 12 kpc, with values for $\Sigma_0$, $r_\mu$, and $\sigma$ taken from Table 1, as described below. An exponential surface density curve was also fitted (dashed line) with $r_0 = r_\mu$ ($r_0$ is the galaxy scale length) and a free value for $\Sigma_0$ and this is overlain for comparison. The corresponding surface density distributions are shown in Figure 2.

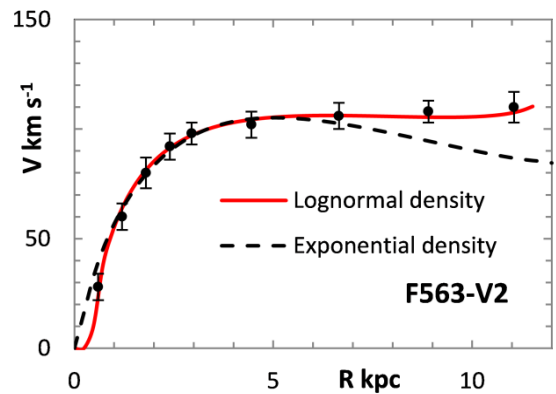

Figure 1. Fitted rotation curves for F563-V2 with log-normal (solid line) and exponential (dashed line) density distributions. Rotation curve data from de Blok, Walter & Brinks et al. (2008).

The exponential RC rises more steeply at small radius, and falls more rapidly at large radius, while the log-normal RC has a later, but steeper initial rise and a long flattened tail leading to differing theoretical disk masses, being $2.19 \times 10^{10}$ and $1.54 \times 10^{10}$ $M_\odot$ for the log-normal and exponential curves respectively. One point immediately evident is the truncated cut-off to the log-normal curve at $R_{max}$ leading to a terminal upturn in the RC, as discussed later. Interestingly, the freely fitted parameter for $r_\mu$ (2.35 kpc) is close to the scale length of the (outer) disk found by Herrmann, Hunter & Elmegreen (2013) of 2.16±0.1 kpc for F563-V2 in the B band. However, the (log) disk mass for F563-V2 is higher than the total disk mass quoted by McGaugh (2012) (10.34±0.15 and 9.83±0.2 respectively) by a factor of dex0.51±0.25.

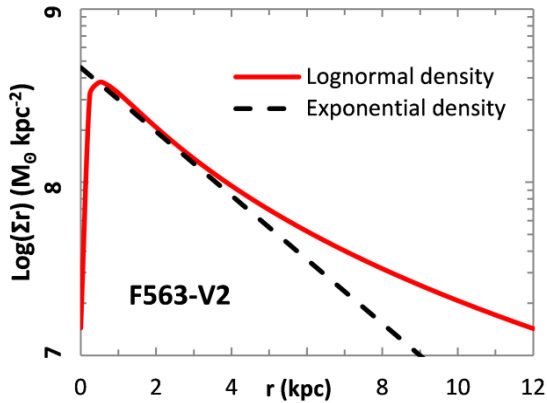

Figure 2. Log-normal surface density plot for F563-V2 (solid line) with modelled exponentially decreasing surface density (dashed line) overlain, to generate the RCs of Figure 1.

The best fit for a log-normal density distribution to the published curve was generated using a bespoke algorithm designed to modify the three free parameters through iterative feedback, using weighted regression analysis to minimise the linear least squares errors for each data point, inversely weighted to the quoted velocity error bars. The derived gravitational dynamic disk mass was computed from the integral of the density curve using Equation 3, but no attempt was made to quantify this in terms of the stellar mass, gas, HI or other components.

The majority of the curves did not require a bulge component to yield a good fit, but where a bulge was added, generally to accommodate an initial peak such as in NGC 6946, NGC 7793, or UGC 2885, or a kink such as DDO 154, this was a uniform-density spherical bulge terminating at $R_{bulge}$, as listed in Table 1, and the bulge mass did not materially affect the overall curve fitting at larger radii. The velocity errors are those quoted in the source documents. The mass errors were computed from the mass variance generated from computed velocity curves for the two-sided 95% confidence intervals for the data points in each data set.

The fitting algorithm was run for 37 representative galaxies for which good RCs have been published (Table 1), which also lists the three free parameters ($\Sigma_0$, $r_\mu$ and $\sigma$) for the log-normal curves, and these results are presented in Figures 1 and 3-38. It may be noted that a variety of velocity curves can be generated from this simple model, even though the log-normal surface density curves look superficially similar, differing only in the height, position and width of the peak (corresponding approximately to the three free parameters). The resulting velocity curves show three broad types: those that continue to rise towards $R_{max}$; those that rise to a flat plateau; and those that rise to a maximum before declining again, in agreement with Verheijen's classification of 'R', 'F' and 'D' type curves respectively (Verheijen 2001).

**Table 1.** Galaxies modelled with a log-normal disk density distribution, associated parameters, observational peak rotation velocity, and theoretically derived total disk mass.

| Galaxy | Ref | Rmax (kpc) | $r_\mu$ (kpc) | $\sigma$ | $\Sigma_0$ (M$_\odot$ kpc$^{-2}$) | Rbulge (kpc) | Mbulge (log M$_\odot$) | Mdisk (log M$_\odot$) | Vmax (km s$^{-1}$) | Type (13) |
|---|---|---|---|---|---|---|---|---|---|---|
| F563-V2 | 2 | 12.0 | 2.35 | 1.22 | 1.80E+08 | | | 10.34 ± 0.15 | 113 ± 5 | F |
| F568-1 | 2 | 15.0 | 3.90 | 1.25 | 1.57E+08 | | | 10.65 ± 0.15 | 139 ± 5 | F |
| F568-3 | 2 | 14.0 | 5.00 | 1.18 | 7.55E+07 | | | 10.43 ± 0.15 | 108 ± 5 | F |
| F568-V1 | 2 | 19.0 | 3.85 | 1.37 | 1.24E+08 | | | 10.63 ± 0.15 | 124 ± 5 | F |
| F574-1 | 2 | 16.0 | 5.07 | 1.38 | 6.71E+07 | 0.20 | 7.13 | 10.45 ± 0.15 | 108 ± 5 | F |
| DDO 154 | 1 | 8.5 | 3.01 | 1.19 | 2.58E+07 | 0.37 | 7.53 | 9.53 ± 0.15 | 56 ± 15 | R |
| M31 | 3 | 34.8 | 4.28 | 1.15 | 5.85E+08 | | | 11.45 ± 0.15 | 255 ± 12 | D |
| Milky Way | 4 | 21.0 | 5.20 | 1.90 | 4.64E+08 | 0.33 | 9.77 | 11.44 ± 0.15 | 298 ± 20 | D |
| NGC 925 | 1 | 15.0 | 10.14 | 1.39 | 4.58E+07 | 0.33 | 7.73 | 10.56 ± 0.18 | 123 ± 10 | R |
| NGC 1705 | 5 | 4.8 | 4.10 | 1.70 | 4.52E+07 | 0.20 | 6.83 | 9.67 ± 0.18 | 72 ± 5 | R |
| NGC 2403 | 1 | 20.0 | 4.29 | 1.40 | 1.54E+08 | | | 10.80 ± 0.15 | 142 ± 10 | F |
| NGC 2683 | 6 | 20.0 | 1.26 | 1.27 | 1.32E+09 | 0.25 | 8.82 | 10.89 ± 0.15 | 211 ± 12 | D |
| NGC 2841 | 1 | 35.5 | 2.10 | 1.70 | 1.86E+09 | | | 11.65 ± 0.15 | 321 ± 10 | D |
| NGC 2903 | 1 | 32.0 | 2.50 | 1.55 | 6.80E+08 | | | 11.27 ± 0.15 | 212 ± 5 | D |
| NGC 2915 | 5 | 16.0 | 2.83 | 1.63 | 9.07E+07 | | | 10.30 ± 0.15 | 93 ± 10 | F |
| NGC 2976 | 1 | 3.0 | 2.50 | 1.60 | 9.60E+07 | | | 9.57 ± 0.15 | 90 ± 15 | R |
| NGC 3198 | 7 | 33.0 | 3.80 | 1.59 | 2.34E+08 | | | 11.09 ± 0.15 | 154 ± 8 | F |
| NGC 3521 | 1 | 31.5 | 2.50 | 1.40 | 8.40E+08 | | | 11.31 ± 0.15 | 235 ± 10 | D |
| NGC 3726 | 8 | 32.0 | 5.94 | 1.40 | 1.65E+08 | | | 11.16 ± 0.15 | 169 ± 15 | F |
| NGC 3741 | 9 | 7.0 | 2.52 | 1.56 | 2.74E+07 | | | 9.42 ± 0.18 | 48 ± 15 | R |
| NGC 4217 | 8 | 16.0 | 2.97 | 0.97 | 4.77E+08 | 0.50 | 8.87 | 10.90 ± 0.15 | 193 ± 10 | D |
| NGC 4389 | 8 | 5.0 | 3.70 | 1.28 | 9.58E+07 | 0.35 | 7.66 | 9.95 ± 0.10 | 115 ± 2 | R |
| NGC 6946 | 1 | 19.5 | 3.67 | 1.13 | 2.91E+08 | 0.40 | 9.17 | 10.92 ± 0.15 | 170 ± 2 | D |
| NGC 7331 | 1 | 26.0 | 2.79 | 1.45 | 9.33E+08 | | | 11.40 ± 0.15 | 262 ± 10 | D |
| NGC 7793 | 1 | 8.0 | 2.57 | 0.67 | 1.75E+08 | 0.69 | 9.05 | 10.11 ± 0.16 | 111 ± 10 | D |
| UGC 128 | 10 | 50.0 | 11.32 | 1.28 | 5.45E+07 | 0.90 | 8.79 | 11.16 ± 0.15 | 138 ± 5 | F |
| UGC 2885 | 11 | 130.0 | 16.20 | 2.44 | 1.79E+08 | 0.70 | 10.33 | 12.30 ± 0.15 | 310 ± 12 | F |
| UGC 6399 | 8 | 8.5 | 3.76 | 1.35 | 7.31E+07 | | | 10.10 ± 0.15 | 93 ± 5 | R |
| UGC 6446 | 8 | 15.5 | 3.83 | 1.70 | 6.05E+07 | | | 10.28 ± 0.15 | 87 ± 12 | F |
| UGC 6667 | 8 | 8.5 | 4.02 | 1.46 | 6.35E+07 | | | 10.07 ± 0.18 | 90 ± 10 | R |
| UGC 6818 | 8 | 7.3 | 5.66 | 1.46 | 3.29E+07 | | | 9.84 ± 0.18 | 77 ± 15 | R |
| UGC 6917 | 8 | 11.0 | 4.17 | 1.46 | 9.62E+07 | | | 10.38 ± 0.15 | 113 ± 5 | R |
| UGC 6923 | 12 | 9.0 | 3.55 | 1.55 | 6.78E+07 | | | 10.07 ± 0.15 | 96 ± 5 | R |
| UGC 6969 | 12 | 8.0 | 5.94 | 1.39 | 3.59E+07 | | | 9.94 ± 0.18 | 89 ± 8 | R |
| UGC 6973 | 8 | 7.3 | 0.96 | 1.76 | 1.17E+09 | | | 10.58 ± 0.18 | 184 ± 10 | D |
| UGC 6983 | 8 | 15.0 | 3.06 | 1.21 | 1.44E+08 | | | 10.46 ± 0.15 | 112 ± 8 | F |
| UGC 7089 | 8 | 9.0 | 4.75 | 1.50 | 4.00E+07 | | | 9.97 ± 0.18 | 86 ± 12 | R |

(1) de Blok et al. (2008); (2) Swaters, Madore & Trewhella (2000); (3) Carignan et al. (2006), Rubin & Kent Ford (1970); (4) Bhattacharjee, Chaudhury & Kundu (2014); (5) Elson, de Blok & Kraan-Korteweg (2012); (6) Casertano & van Gorkom (1991); (7) van Albada, Bahcall, Begeman & Sancisi (1985); (8) Sanders & Verheijen (1998); (9) Begum, Chengalur & Karachentsev (2005); (10) de Blok & McGaugh (1998); (11) Roelfsema & Allen (1985); (12) Bottema (2002); (13) Verheijen (2001) type classification: 'R' R-type; 'F' F-type; 'D' D-type.

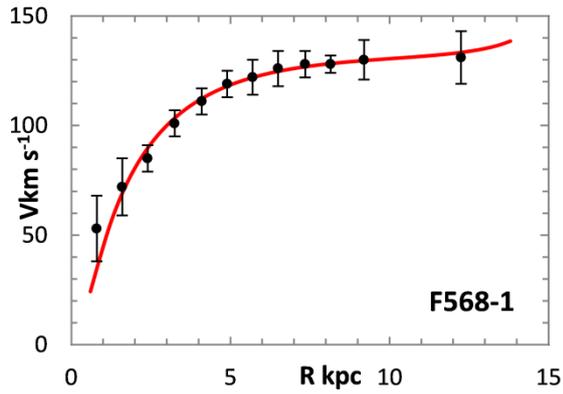
Fig 3. Log-normal rotation curve for F568-1.

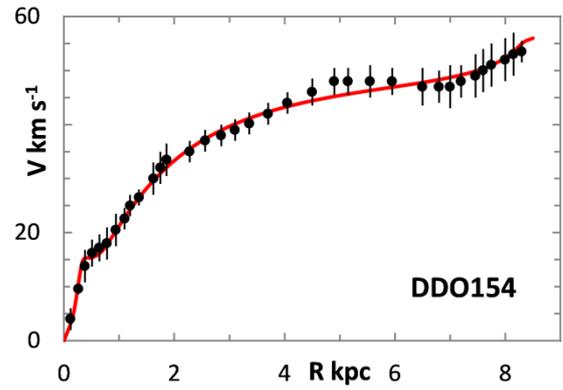
Fig 7. Log-normal rotation curve for DDO 154.

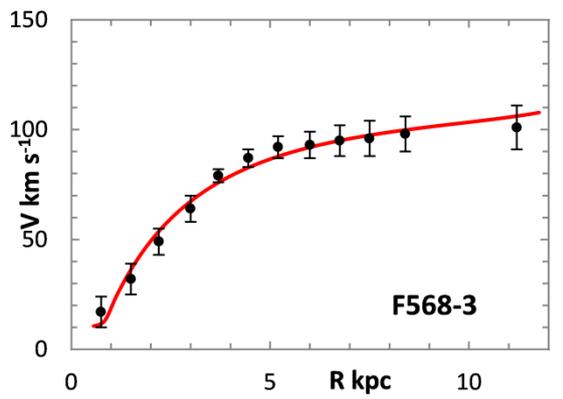
Fig 4. Log-normal rotation curve for F568-3.

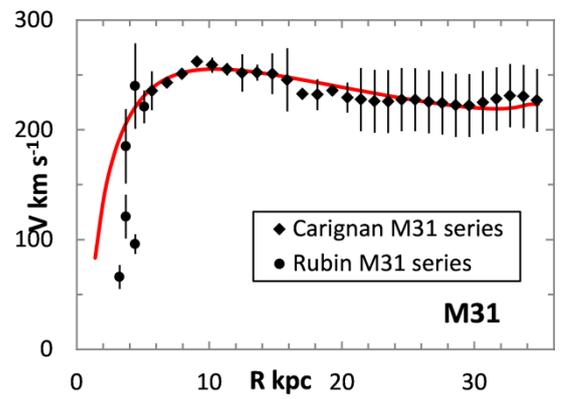
Fig. 8. Log-normal rotation curve for M31.

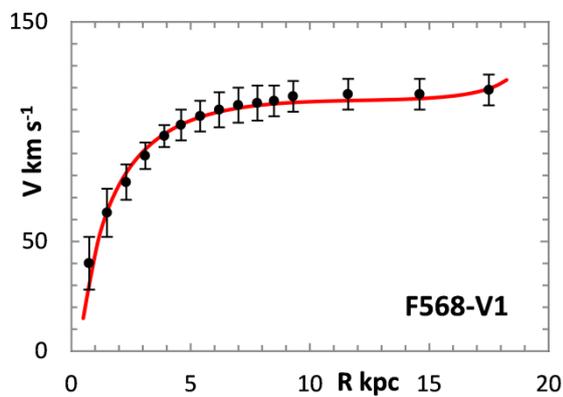
Fig 5. Log-normal rotation curve for F568-V1.

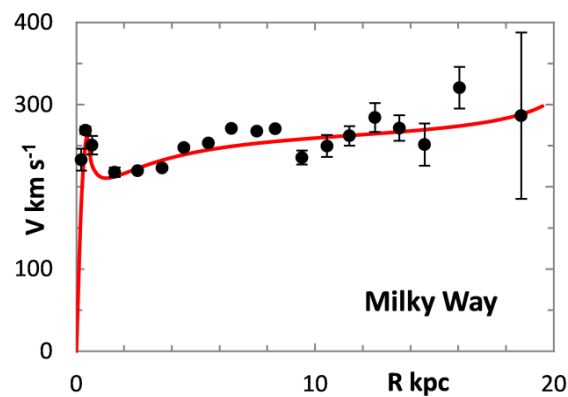
Fig 9. Log-normal rotation curve for Milky Way.

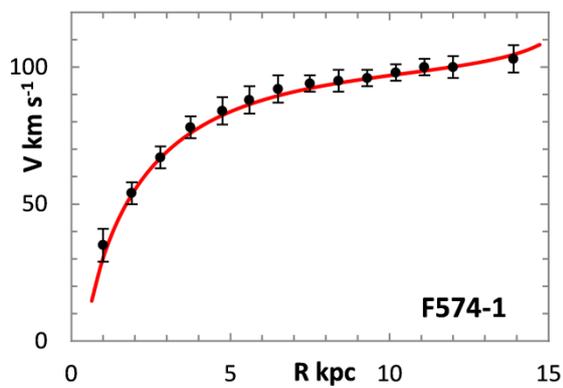
Fig. 6. Log-normal rotation curve for F574-1.

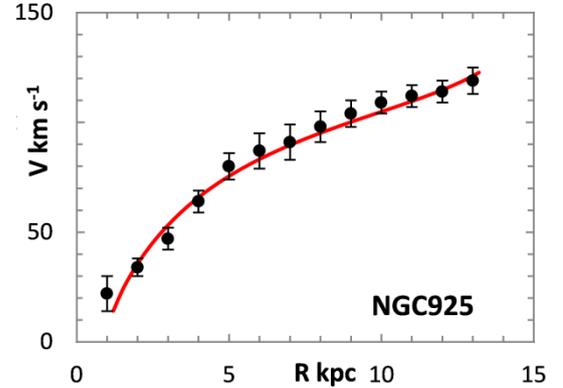
Fig. 10. Log-normal rotation curve for NGC 925.

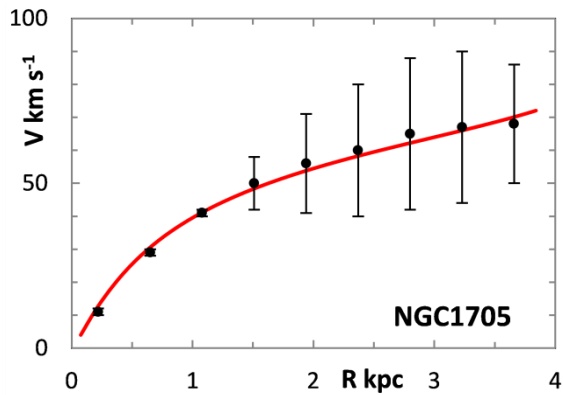
Fig 11. Log-normal rotation curve for NGC 1705.

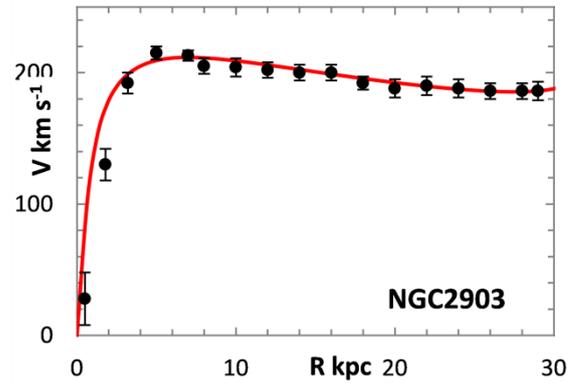
Fig 15. Log-normal rotation curve for NGC 2903.

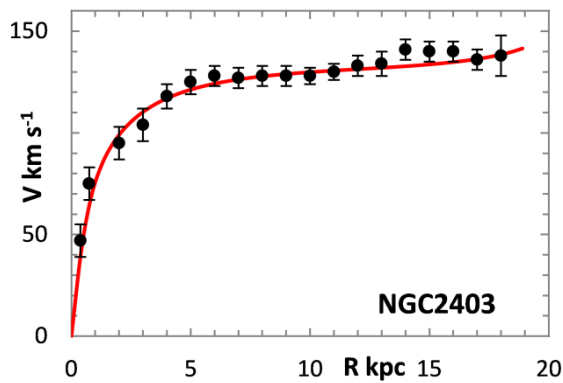
Fig 12. Log-normal rotation curve for NGC 2403.

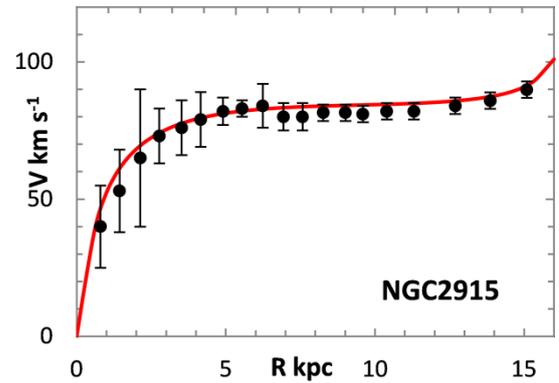
Fig 16. Log-normal rotation curve for NGC 2915.

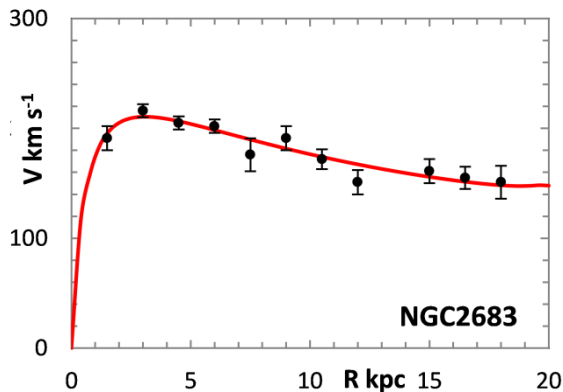
Fig 13. Log-normal rotation curve for NGC 2683.

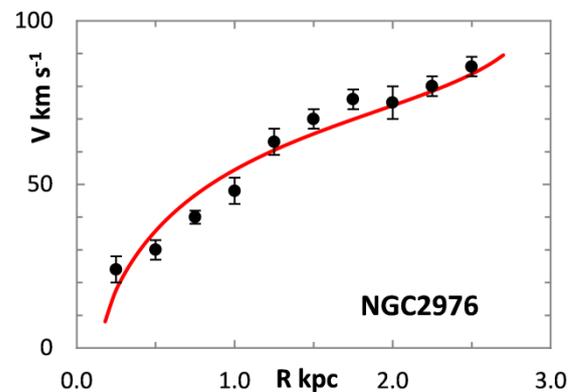
Fig 17. Log-normal rotation curve for NGC 2976.

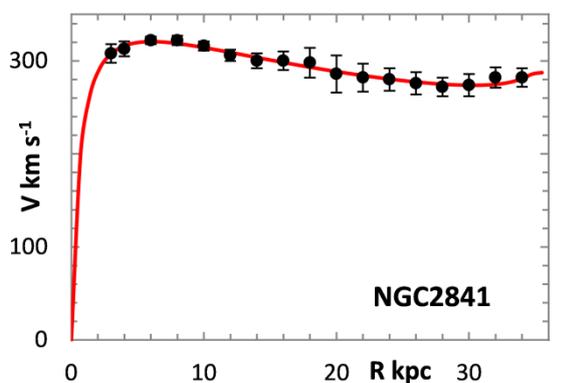
Fig 14. Log-normal rotation curve for NGC 2841.

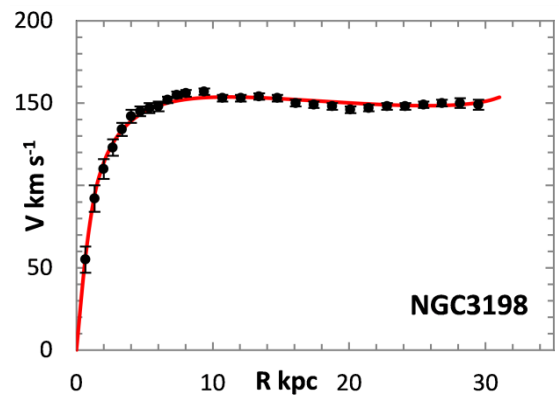
Fig 18. Log-normal rotation curve for NGC 3198

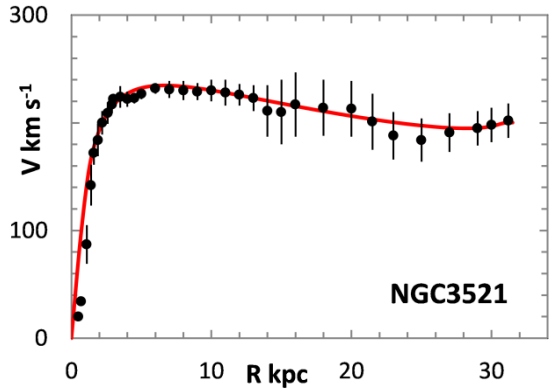
Fig 19. Log-normal rotation curve for NGC 3521.

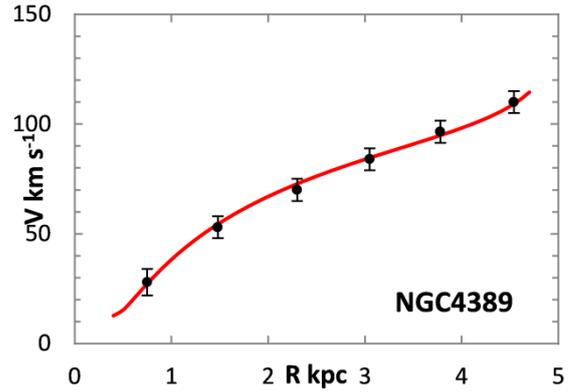
Fig 23. Log-normal rotation curve for NGC 4389.

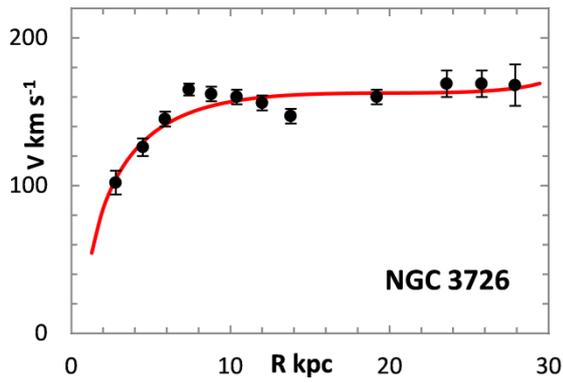
Fig 20. Log-normal rotation curve for NGC 3726.

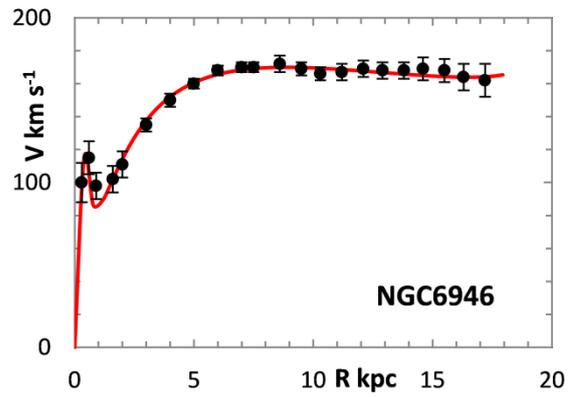
Fig 24. Log-normal rotation curve for NGC 6946.

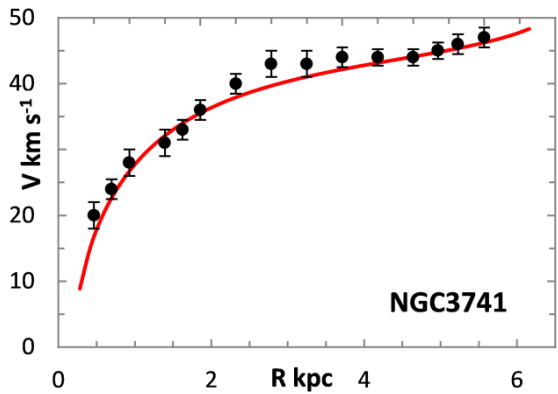
Fig. 21. Log-normal rotation curve for NGC 3741.

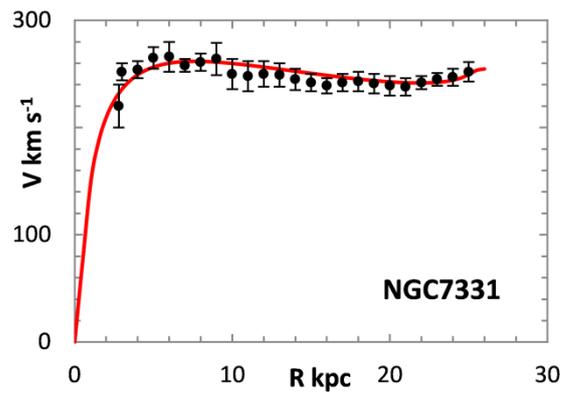
Fig 25. Log-normal rotation curve for NGC 7331.

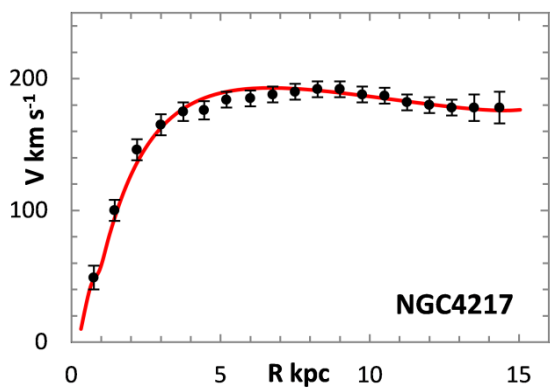
Fig 22. Log-normal rotation curve for NGC 4217.

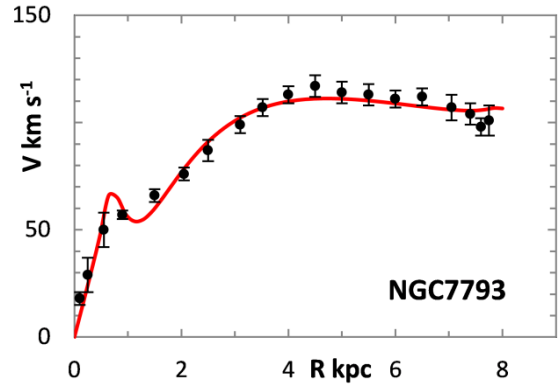
Fig 26. Log-normal rotation curve for NGC 7793.

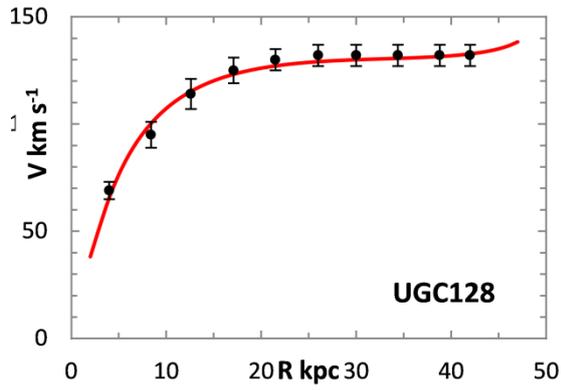
Fig 27. Log-normal rotation curve for UGC 128.

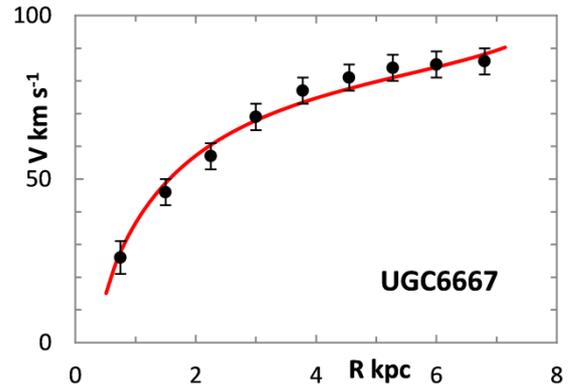
Fig 31. Log-normal rotation curve for UCG6667.

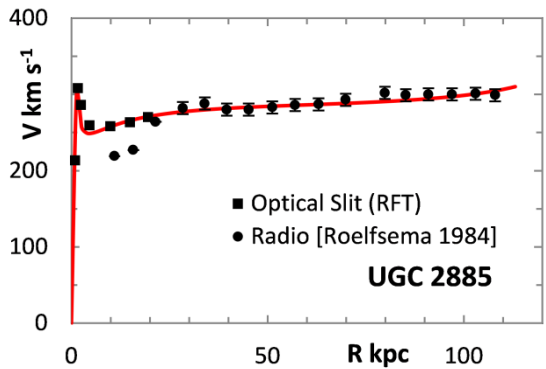
Fig 28. Log-normal rotation curve for UGC 2885.

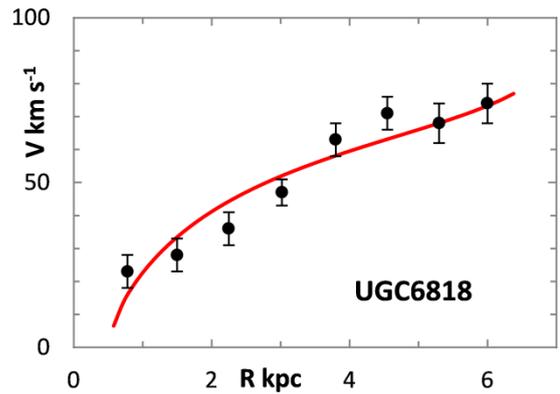
Fig 32. Log-normal rotation curve for UGC 6818.

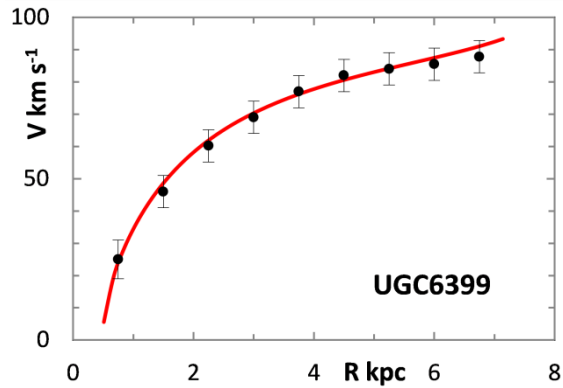
Fig 29. Log-normal rotation curve for UGC 6399.

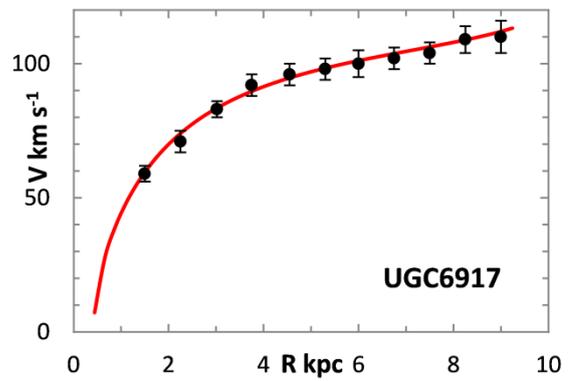
Fig 33. Log-normal rotation curve for UGC 6917.

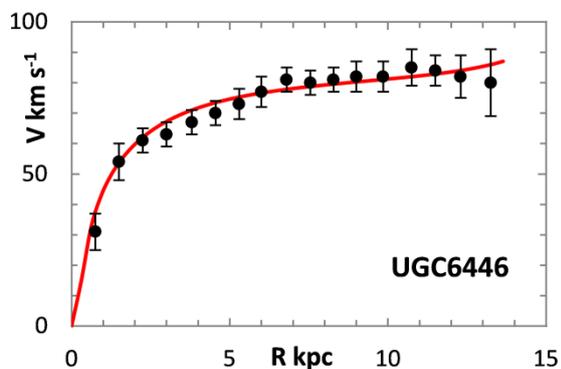
Fig 30. Log-normal rotation curve for UGC 6446.

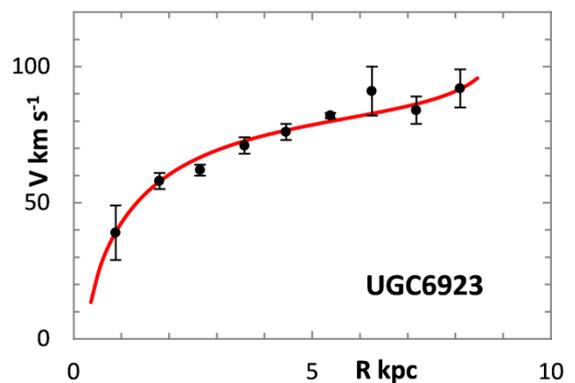
Fig 34. Log-normal rotation curve for UGC 6923.

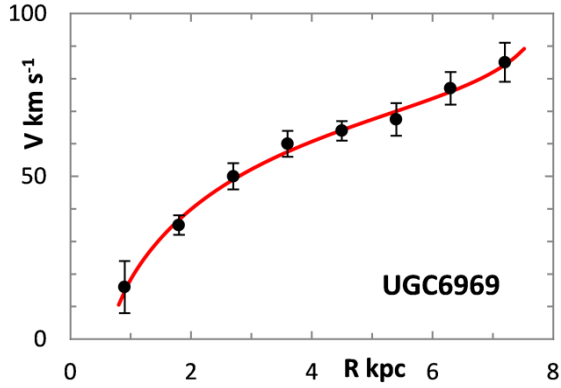

Fig 35. Log-normal rotation curve for UGC 6969.

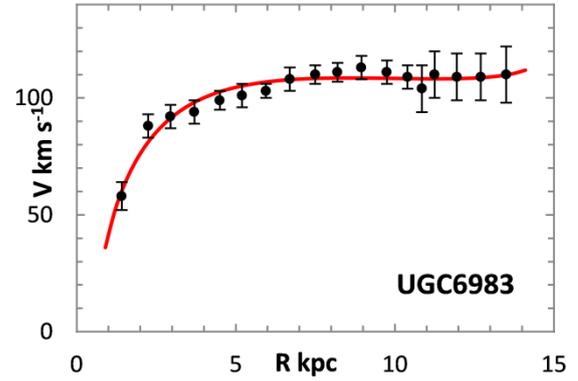

Fig 37. Log-normal rotation curve for UGC 6983.

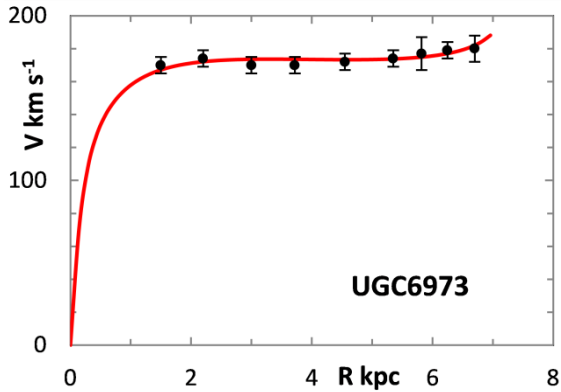

Fig 36. Log-normal rotation curve for UGC 6973.

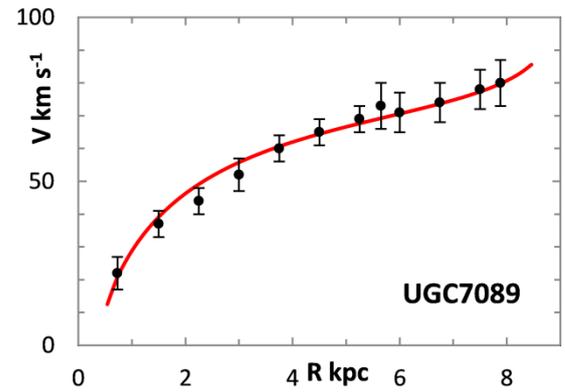

Fig 38. Log-normal rotation curve for UGC 7089.

One problem for galaxy disk models is to select an appropriate cut off radius, $R_{max}$, which defines the total disk mass in terms of a disk boundary. An analysis of this has been presented by Bizyaev and Zasov (2002) who found the azimuthally averaged brightness of the stellar disk in spiral galaxies decreased with galactocentric distance beyond a certain radius (3–5 inner-disk scale lengths), obeying in most cases an exponential law, before becoming generally truncated with the radial surface brightness profile steepening sharply. More recently, Erwin, Beckman and Pohlen (2005) presented the radial brightness profiles of a number of barred S0–Sb galaxies with "anti-truncations", with outer profiles distinctly shallower in slope than the main disk profile, later classifying the profiles outside the bar region into three main groups: Type I (single-exponential), Type II (down-bending), and Type III (up-bending) (Erwin et al 2008).

Herrmann, Hunter and Elmegreen (2013) examined the stellar disk profiles of 141 dwarf galaxies, and fitted single, double, or even triple exponential profiles in up to 11 passbands. Using a simple exponential disc plus stellar halo model, based on current observational constraints, Martin-Navarro et al (2014) showed that truncations in face-on projections occurred at surface brightness levels comparable to the brightness of stellar haloes at the same radial distance. They suggested that stellar haloes outshine the galaxy disc at the expected position of the truncations, allowing these to be studied only in highly inclined (edge-on) orientations.

Although the analyses of the brightness profiles out to extremely faint isophotes suggests in many cases that the profile steepens abruptly at some large distance, defined as the disk cutoff radius $R_{max}$, this cutoff does not imply a total absence of gas and stars at large galactocentric distances. The HI profile width is often taken as a standard diameter for galactic measurements (Singhal 2008), but in a number of cases, rotating HI disks extend far beyond the optical boundary, and isolated star-forming regions are sometimes seen very far from the galactic centres. Bizyaev and Zasov (2002) commented that the radial velocity dispersion of the stars in the disk (10–20 km/s for regions near the disk periphery) "smears out" the edges, so the cutoff cannot be completely sharp, and the uncertainty in $R_{max}$ estimates can exceed 1–2 kpc.

Different tracers of the rotation curve can probe different radial regimes of the RC, and HI data tend to extend much further than optical (e.g., H-alpha) data, again confounding the most appropriate value for $R_{max}$. Observations of relatively nearby galaxies show that stars are still born in places where the observed gas density is below the critical threshold, albeit at lower rates, therefore faint extensions of stellar disks may exist even at $R > R_{max}$ (e.g. in M33) (Bizyaev & Zasov 2002). For the model in this paper, $R_{max}$ was generally chosen to be one bin size beyond the last recorded value for the rotation velocities of the selected galaxies.

The difficulty in defining rotation velocity for the bTFR has also been well documented (e.g. Verheijen 2001), and several definitions have been used such as Peak Velocity, Maximum Velocity, Flat Velocity, Terminal Velocity, and Velocity at percent of $R_{max}$ (Sofue & Rubin 2001; McGough 2012).

The present analysis uses the peak velocity to derive the bTFR plots associated with the reported velocity curves. This generally coincides with the flat velocity of McGaugh (2011) except for "R" type curves, where $V_{rot}$ is still strongly increasing at the maximum radial measurement, and some "D" type curves which have an early peak velocity. For the later, $V_{max}$ was taken at the peak velocity, which may be well inside the disk. For "F" type curves, $V_{max}$ was generally taken to be the maximum value of $V_{rot}$ provided by the literature source. For "R" type galaxies, $V_{rot}$ is still rising at the last reported radius. Extending $R_{max}$ to 1 bin beyond the last reading may artificially increase $V_{max}$, especially in conjunction with the truncation effect at $R_{max}$. In these cases, $V_{max}$ was generally taken to lie between the last "official" reading, and the value predicted by the model at $R_{max}$.

## 3. THE TULLY-FISHER RELATION FOR THE LOG-NORMAL DISK MASSES

In their classic paper, Tully and Fisher (1977) proposed a convenient empirical relation between luminosity and line width (the Tully-Fisher relation, or TFR). In this context, luminosity is a proxy for stellar mass, which in turn depends on the total mass, and the physical basis of the TFR is widely presumed to be a relation between a galaxy's total mass and rotation velocity (e.g., Freeman 1999). Much subsequent work has involved attempts to quantify the disk mass in terms of the luminosity, for example by assuming an appropriate initial mass function (IMF) such as that of Salpeter (1955). However, Freeman (1999), McGaugh et al (2000), and McGaugh (2012) have demonstrated that luminosity is not a perfect predictor of mass, as the stellar mass-to-light ratio can vary with galaxy type. Consequently, the TFR may have different slopes depending on the luminosity bandpass (e.g., Tully et al. 1998; Bell & de Jong 2001; Verheijen 2001; Courteau et al. 2007; Noordermeer & Verheijen 2007), and bright galaxies tend to lie above the extrapolation for low-luminosity galaxies (Persic & Salucci 1991; Matthews et al. 1998). Masters, Springob & Huchra (2008) found that the slope of the TFR became steeper as the wavelength increased, being close to $L \propto v^4$ in the K band and $L \propto v^{3.6}$ in the J and H bands. In all three bands the relation was steeper for later-type spirals. McGaugh et al. (2000) found that a more fundamental relationship between the baryonic mass and rotation velocity does indeed exist, provided that both stellar and gas mass are considered (Milgrom & Braun 1988; McGaugh 2005).

Although originally proposed as a function of luminosity, the bTFR appears to be a fundamental relationship that is linear (in log space) over many decades in mass (Verheijen 2001; Gurovich et al. 2004; McGaugh 2005; Pfenniger & Revaz 2005; Begum et al. 2008; Stark et al. 2009; Trachternach et al. 2009; McGaugh 2012), and should not distinguish between stellar mass and mass in other forms.

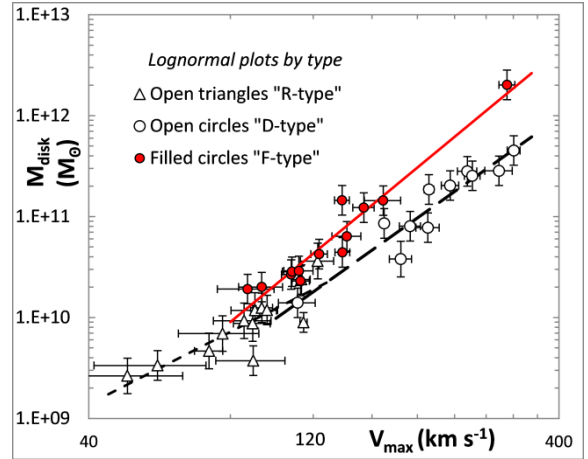

Figure 39. Slopes for 3 galaxy types: "F" (solid line), "D" (long dashed line) and "R" (short dashed line).

The 37 galaxies presented in this paper are plotted in Figure 39, using the theoretical total galactic masses derived from the log-normal model against $V_{max}$ divided into the three types of Verheijen (2001): "R" type (open triangles), "D" type (open circles) and "F" type (filled circles). It will be noted that the three groups appear to be distinct, and with different slopes. The "F" type have a slope of 3.84±0.3, the "D" type a slope of 3.37±0.39, while the "R" type have a shallower slope of 2.38±0.45. Galaxies with declining RCs were found to lie systematically on the high-velocity side to those with flat velocities, in conformation to the findings of Noordermeer & Verheijen (2007), who also suggested that there may be a change in slope in the bTFR at the high-luminosity end.

Pfenniger & Revaz (2005) suggested that the galactic baryonic mass is likely to consist not only of the detected baryons, stars and gas, but also of a dark baryonic component proportional to the HI gas, and the bTFR can be substantially improved when the HI mass is multiplied by a factor of about 3, reinforcing the suggestion made in several works (Bosma 1981; Hoekstra, van Albada & Sancisi 2001) that mass within galactic disks must be a multiple of the HI mass, and that galactic disks are substantially, if not necessarily fully, self-gravitating. Gurovich et al. (2010) considered that a larger fraction of ionized undetected baryons is required in the more massive galaxies to steepen the slope of the theoretical bTFR to its observed value, and that ionized (warm) gas in the more massive galaxies (e.g., Maller & Bullock 2004; Fukugita & Peebles 2006) may turn out to be more significant in this respect.

For dense, bright galaxies, $M_{total}$ is generally represented by $M_{stars}$ (M*), but this is prone to errors as described by McGaugh (2011), who suggested that taking the mass of gas-rich low surface brightness (LSB) galaxies might provide a more accurate estimate for the total galactic mass. The 37 galaxies presented in this paper are also shown in Fig 40, overlying the plots from McGaugh's (2011) figure. The solid line is the RMS mean slope of the combined galaxies, with a value of 4.02±0.14. However, the computed log-normal masses have an intercept of 76±35 $M_\odot$ km$^{-4}$ s$^4$, compared to an intercept of 47±6 for the projected observational masses of McGaugh (2012).

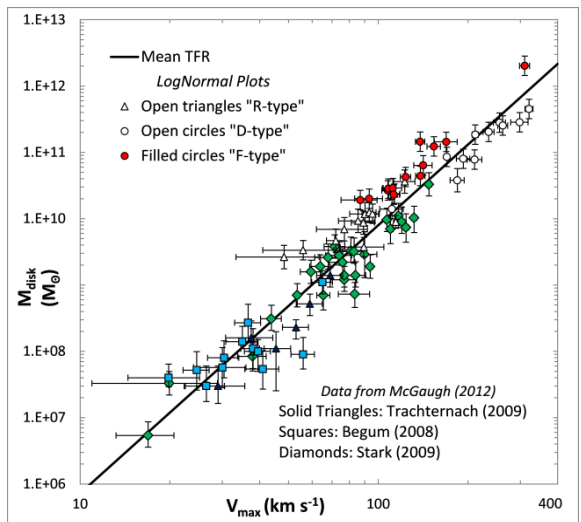

Figure 40. Log-Normal bTFR plots (open and filled circles, open triangles) and the LSB galaxy plots (adapted from McGaugh 2012) with the combined mean slope.

Nine galaxies common to both the log-normal plots and McGaugh's plots show only modest correspondence, with a mean log(mass) of 10.40±0.16 for the log-normal and 9.79±0.21 for the McGaugh galaxies (McGaugh 2012), with a mean excess in total gravitational mass of dex0.61±0.26 over McGaugh's baryonic mass. Therefore, for these nine galaxies, the assumption of a log-normal disk mass-density distribution predicts that the amount of gravitational matter assumed to be in the disk (in any form, baryonic or otherwise) is on average four times as much as the baryonic masses calculated by McGaugh (2012). Interestingly, van Albada & Sancisi (1986) have previously noted that mass models of spiral galaxies are able to reproduce the observed rotation curves in the inner regions, but fail to do so increasingly towards and beyond the edge of the visible material. They found that the discrepancy in the outer region could be accounted for by invoking dark matter, with some galaxies requiring at least four times as much dark matter as luminous matter.

## 4. DISCUSSION

The motion of a test mass in the field of a thin, massive gravitational disk is exquisitely sensitive to the surface density profile of the disk and its termination profile. The generation of the 37 galaxy velocity profiles presented in this paper assumed that most of the galactic mass is in the disk, and gravity is Newtonian. A best-fit algorithm was used to generate the curves of Figures 1 and 3 – 38 using a truncated log-normal surface density distribution function. The log-normal models closely matched the shape of observational rotation velocities, and the predicted masses for these curves fitted a baryonic Tully-Fisher relation reasonably well over a wide range of galaxy sizes, from LSB galaxies to massive high-luminosity disks.

The resultant velocity profiles are highly sensitive to the three parameters intrinsic to this function, and three distinct types of velocity curves can be generated by the model, in broad agreement with Verheijen's classification (Verheijen 2001). All three types may show a characteristic terminal increase in velocity associated with a disk mass/density that truncates abruptly, as first shown by Casertano (1983), which also appears to fit some of the observational data, although this would be reduced by using a more gradual edge decay such as the termination profiles of Herrmann et al. (2013). However, because the measurement of rotation velocities requires the presence of observable baryonic mass in the form of stars, HI, gas, etcetera, and the disk mass affects the gravitational field and hence rotation velocities so strongly, it is unlikely that the classical Keplerian decay curve will be observed, except for objects sufficiently remote that the galaxy may be treated as a point mass.

The exponential disk profile has a profound place in astronomy. Surface photometry indicates that most spiral and S0 galaxies have an exponential disk component with radial surface-brightness distribution $I(r) = I_0 e^{-r/r_0}$ which implies a surface density distribution $\Sigma(r) = \Sigma_0 e^{-r/r_0}$, and a spheroidal bulge component whose properties differ widely between galaxies (Freeman 1970). The parameter $r_0$ is the scale length of the galaxy, with dimensions of typically a few kpc (e.g. for the Milky Way, $r_0 \approx 4$ kpc) (Peacock 1999). Though this cannot be equated directly with the parameter $r_\mu$ of the log-normal probability distribution of Equation 2 ($r_\mu \approx 5.2$ kpc for the Milky Way), both have a physical interpretation in the shape of the disk density profile.

Although a number of theoretical predictions propose the existence of DM to address the mass discrepancy problem in the universe and within galaxies, this model suggests it may be sufficient for the distribution of any DM to be confined to the disk profile, possibly to account for any remaining discrepancy between observed baryonic mass and the theoretically required disk mass, as suggested for example by Gurovich et al. (2010) and Jalocha et al. (2010), who considered that at smaller scales, the contribution of non-baryonic DM to spiral galaxy masses could be much less than anticipated in spherical halo models. By considering the observed radial velocities, positions, and distances of stars in the Sagittarius stream, one can put constraints on the shape of the gravitational potential of the Milky Way that show that it cannot be as flattened as the stellar disk (e.g. Law & Majewski, 2010; Vera-Ciro & Helmi, 2013). However, even in the presence of a prominent spheroidal component, the disk contributes the major part of the total light and angular momentum. In M31 for example, >75% of the blue light and >95% of the total angular momentum come from its disk (de Vaucouleurs 1958; Takase 1967), and the velocity profiles of LSB galaxies have been well fitted to an exponential disk profile (Kassin, de Jong & Weiner 2006; McGaugh 2011, private communication). Conversely, the theoretical curves for an inverse-r velocity profile are essentially flat over an extensive range of galaxy sizes (Mestel 1963), and large and massive star-rich galaxies can be fitted to this profile. The log-normal surface density distribution has attributes of both these profiles, with an exponential and an inverse-r component to the density function, but is broadly similar to a true exponential curve over much of the radius (Figure 1), although this equivalence is lost at the extremities of small and large $r$. Despite this similarity, the resultant RCs do show a marked difference between the log-normal and exponential surface density models for "F" type galaxies over the flat portion of the RC as illustrated in Figure 1.

The poorest fits seen in Figures 3 – 38, e.g. the Milky Way, NGC 2976, NGC 3726 and UGC 6818, may link with non-axisymmetric features such as bars or strong spiral arms, as suggested by Jalocha et al. (2010), and thus might be improved by considering more realistic models. Despite these discrepancies, and although unable to accommodate other measurements of the shape of the galaxy gravitational potential, the resultant RCs show a good overall fit to the observational data, and the theoretical total disk masses generated by the log-normal density distribution model can accommodate a scenario in which the total mass distribution is confined to a thin plane without requiring a dark-matter halo or the use of MOND.

**ACKNOWLEDGMENTS**

I would like to thank Stacy McGaugh for discussion on inverse-r and exponential disk models, Anatoly Zasov for comments about disk stability and DM, David Keeports for his encouragement in preparing the paper, and the anonymous referee for many helpful suggestions and comments.